\documentclass{elsart}
\def\HI{H{\,\small I}}
\newcommand{\ltsima} {$\; \buildrel < \over \sim \;$}
\newcommand{\gtsima} {$\; \buildrel > \over \sim \;$}
\newcommand{\msun}{{$M_\odot$}}
\newcommand{\lta} {\lower.5ex\hbox{\ltsima}}
\newcommand{\gta} {\lower.5ex\hbox{\gtsima}}
\usepackage{natbib}
\input{psfig}
\begin{document}
\runauthor{Morganti et al.}
\begin{frontmatter}
\title{Large-scale \HI\ structures and the origin of radio galaxies}
\author[raf.tom]{R. Morganti},
\author[raf.tom]{T. Oosterloo},
\author[clive]{C. Tadhunter},
\author[bjorn]{B. Emonts}

\address[raf.tom]{Netherlands Foundation for Research in Astronomy, Postbus 2,
NL-7990 AA, Dwingeloo, The Netherlands}
\address[clive]{Dep. Physics and Astronomy,
University of Sheffield, Sheffield, S7 3RH, UK}
\address[bjorn]{Kapteyn Astronomical Institute, RuG, Landleven 12, 9747 AD,
Groningen, NL}

\begin{abstract}

We present the first results of a study aimed to detect large \HI\
structures in radio galaxies.  In two of the three cases presented
(Coma~A and B2~0648+27), the detection of a large amount of \HI\
distributed over several tens of kpc  suggests a major merger as origin of the
observed system and allows to infer when this merger must have
occurred.  The situation is less clear for the third object (3C~433).
We propose an evolutionary sequence for one of the radio galaxies
studied.  The implications of this study of neutral hydrogen in nearby
radio galaxies for high-$z$ objects and ``normal'' (i.e. radio-quiet)
elliptical galaxies are also discussed.

\end{abstract}
\end{frontmatter}

\begin{figure}[t]
\centerline{\psfig{figure=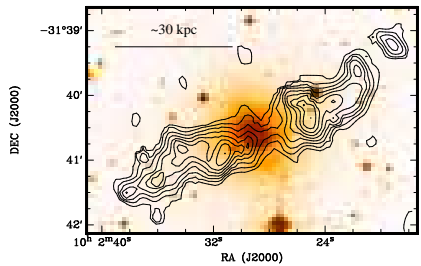,angle=-90,width=6cm}}
\caption{An example of a  large (about 50 kpc diameter), regular \HI\ disk
of more than $10^9 M_{\odot}$ of \HI\ (in contours the total intensity)
around the (radio-quiet) elliptical galaxy NGC~3108 (optical image in
grey scale). See \cite{To1} for more details.}
\label{f1}
\end{figure}

\section{Why neutral hydrogen?}

The study of the formation and the evolution of massive ellipticals is
attracting, at present, a lot of attention. Among these systems, an
interesting sub-class is represented by those hosting powerful radio
sources. The presence of such activity is telling us that the galaxy is
in a particular phase in its evolution. Indeed, the increasing evidence
given that super-massive black holes reside in the centres of many
elliptical galaxies suggests that the period of activity, turning a
``normal'' elliptical in a radio galaxy, could be a -- relatively
short -- phase in the life of many (or all?) ellipticals. When in the
life of the host galaxy this phase takes place and how it is related to
the actual formation of the host galaxy and to its evolution, are 
questions that are still unanswered.
\\
The importance of mergers/interactions in the formation and triggering
of radio galaxies has been brought forward already a long time ago. The
peculiarities often observed in the optical morphology of radio
galaxies (e.g.\ tails, bridges, shells and dust-lanes, see \cite {He1})
is one of the main arguments that has led to this conclusion. \\
However, new results have been recently obtained from the study of the
optical continuum in radio galaxies (with the main goal of
investigating the nature of the UV excess). It has been found
\cite{Ta1, Wi1} that a young stellar population component is present
in at least 30\% of radio galaxies. Interestingly, the galaxies
showing this component are also the most luminous in the far-IR,
indicating a link between an optical starburst and far-IR
activity. These results support the idea of a merger origin for radio
galaxies and suggest that, at least some of them, had an
(ultra-)luminous far-IR galaxy as progenitor \cite{Ta1}. The results
also indicate that the activity starts late after the merger event.
The existence of galaxies where this young stellar component is not
observed indicates that either they are observed at a later stage
and/or they originate from {\sl other type} of mergers.  \\ Thus, the
study of the stellar population appears to provide some important
clues about the origin of radio galaxies.  However, some of the
information derived in this way is still quite uncertain and it is
important to be able to investigate the origin and evolution of these
systems in a complementary way.  This can be done through the study of
{\sl large-scale} structures of neutral hydrogen around radio
galaxies.  Indeed, \HI\ has been successfully used to study the fate
of the gas in a number of cases where interaction/merger is present. The
main advantage of the study of neutral hydrogen is, given the long
timescales involved, its capability of retaining old signatures of the
merger/accretion. Thus, it can be used to derive key information about
the {\sl age, size and type} of the merger and about when  the AGN phase takes
place compared to the starburst phase. 
\\
Neutral hydrogen has recently given interesting results also for
``normal'' (i.e.\ not necessarily radio-loud) elliptical galaxies (see
\cite{To1, To2, S1} and refs therein). Interestingly, a small, but significant, 
fraction, in particular outside clusters of galaxies, contains a sometimes very
large amount of \HI\ (i.e.\ $M_{\rm {HI}} > 10^9$ \msun\footnote{For
$H_\circ = 50$ km s$^{-1}$ Mpc$^{-1}$ and $q = 0.0$}, $M_{\rm
{HI}}/L_B \sim$ 0.02) distributed in very extended (tens of kpc in
size, see e.g. Fig.~1), and often kinematically regular, structures (see e.g.\
\cite{Mo1, To1}).  These structures are long-lived (no major star-formation
is going on) and, because of their large amount of \HI\ and the large
size, they are believed to originate from major mergers between two
disk galaxies.  In addition to this, the often observed decoupling
between the gas and stellar content and kinematics is taken as a
further indication of the external origin of the \HI\ in ellipticals
\cite{Kn1}.  One obvious question is whether there is any
connection between radio galaxies, these gas-rich ``normal''
elliptical galaxies and other major mergers (like the ultra-luminous
far-IR galaxies, ULIRGs).

\begin{figure*}
\centerline{\psfig{figure=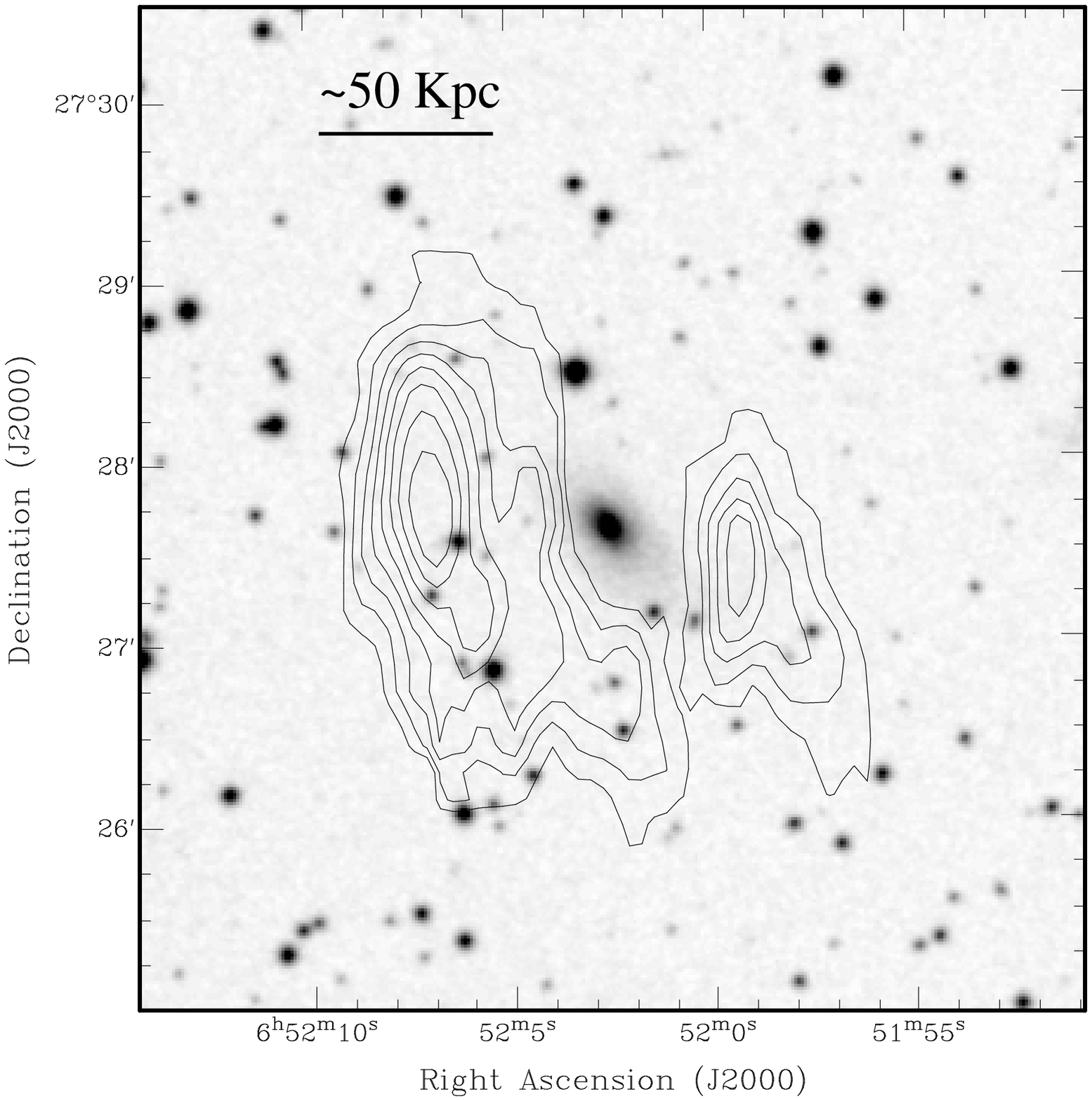,angle=-90,width=10cm}
\psfig{figure=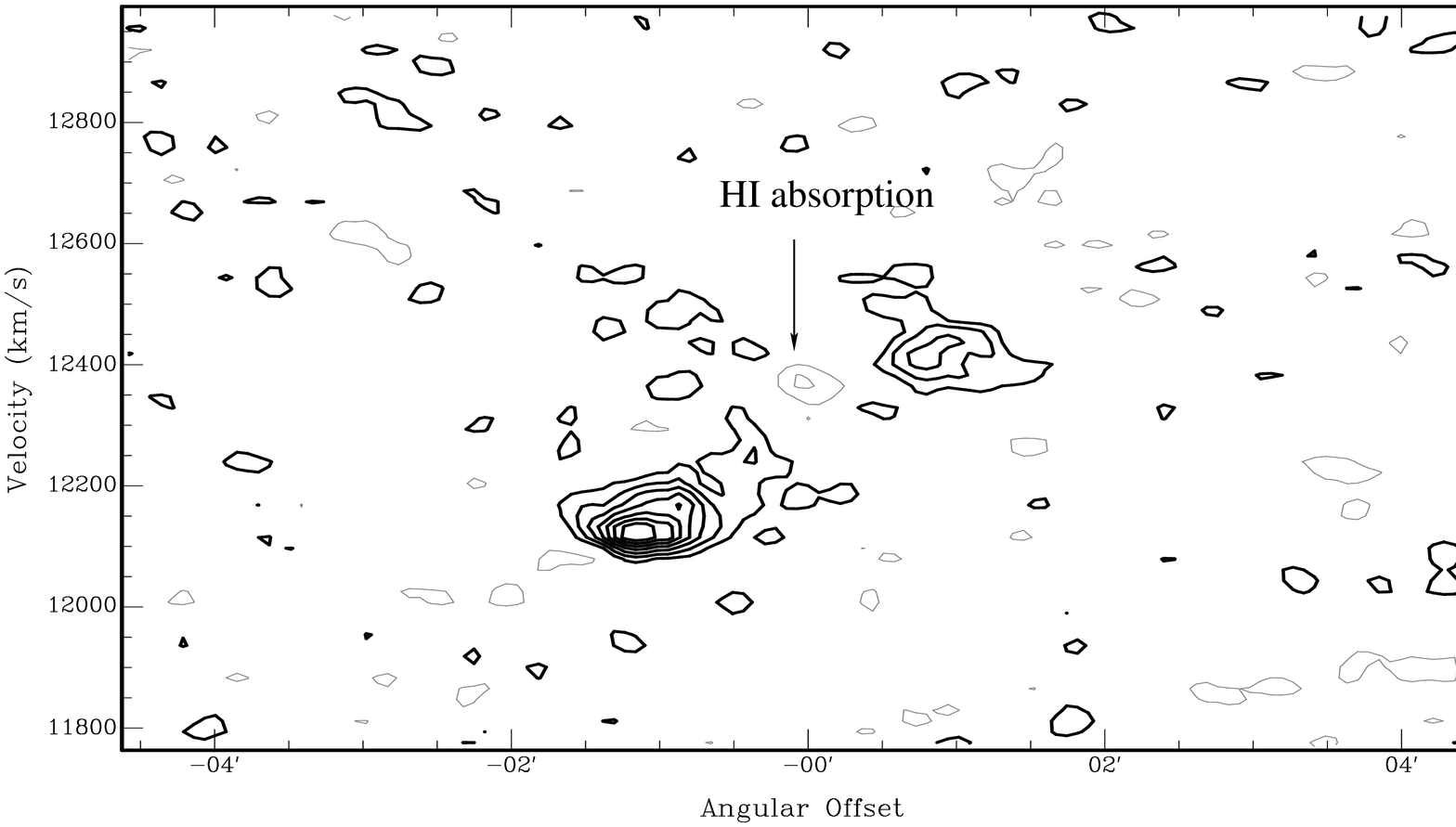,angle=-90,width=7cm}}
\caption{({\sl Left}) \HI\ total intensity contours  of
the radio galaxy B2 0648+27  superimposed on to an optical image.
{\sl (Right)} Position-velocity plot along the major axis. The \HI\ absorption
detected against the unresolved continuum is also indicated.}
\label{f1}
\end{figure*}

\section{Large-scale \HI\ structures in radio galaxies: new results}

For all these reasons, we have begun a study of {\sl large-scale
\HI\ structures} (tens of kpc in size) in and around radio galaxies. 
The presence of neutral hydrogen around radio galaxies was known only
for a handful of objects. With this study we plan not only to increase
the known number of cases, but also to reach a more complete
statistics of its occurrence in order to compare with the situation
for ``normal'' ellipticals. \\ In our search for large-scale \HI\
structures in radio galaxies, we have already at least three
interesting new detections (Coma~A, B2~0648+27 and 3C~433).  This is
quite remarkable given that so far only two radio galaxies of
comparable radio power (Centaurus A and PKS B1718-649, see \cite{Mo3}
for an overview) were known to have large-scale \HI. For the first
time, some of the large-scale \HI\ structures are detected in
absorption against the extended radio lobes.

In the radio galaxy Coma~A, $10^9$ \msun\ of \HI\ was detected in a
disk-like structure of at least 60 kpc in diameter in absorption
against the radio lobes (see \cite{Mo2}). The structure of the gas
disk (and the amount of \HI) suggests that a merger occurred,
involving at least one large gas-rich galaxy, at least a few times
$10^8$ yr ago.  Another example is the nearby radio galaxy B2~0648+27
where \HI\ both in emission and in absorption has been detected
\cite{Mo3}. In emission, we detect a very large amount of \HI\
($M_{\rm HI} = 1.1 \cdot 10^{10}$ $M_\odot$) distributed in a very
extended disk, or ring-like structure, of about 160 kpc in size as
shown in Fig.\ 2 \cite {Mo3}. We also detect \HI\
absorption against the central radio continuum component. The
characteristics of the detected \HI\ and the similarities with some of
the ``normal'' elliptical mentioned  above, are explained as the result
of a major merger event that is likely to have occurred $\lta
10^9$ yr ago.
\\
Extended \HI\ in absorption against the radio lobes has been observed also is
3C~433.  In this radio galaxy (Fig.~3), the preliminary analysis of the data
shows that $\sim 5\times 10^8$ \msun\ of (extended) \HI\ is detected at about
60 kpc from the radio core.  The gas shows a velocity gradient, but at the
moment it is not clear whether the detected \HI\ is part of an extended gas
disk/tail or whether it corresponds to a region of interaction between the ISM
and the radio lobe.  Given the lower amount of \HI\ detected so far in this
object, the origin of this galaxy might be different from the major-merger
suggested in the two previous cases. 
Finally, it is important to notice that both B2~0648+27 and 3C~433 have
a clear young stellar population component visible in their optical
spectra. The final goal will be to compare the information obtained
from the \HI\ data and those obtained from the stellar population.

\section{Can we use the \HI\ to define an evolutionary sequence?}

For B2~0648+27 we have attempted, using the information obtained from
the \HI, to include this object in an evolutionary sequence, partly
already proposed for other gas-rich systems (see e.g.\
\cite{Hi2}). This could provide a possible scenario for the origin of
at least some radio galaxies. \\ The sequence, depicted in Fig.~5,
begins with systems like The Antennae (left) as representative of an
ongoing merger where two galaxies are still identifiable and where a
starburst is occurring.  NGC~7252 (2\( ^{\rm nd}\) from left)
represents a somewhat later stage where the central remnant has
already more or less taken the shape of an elliptical galaxy
\cite{Hi2}.  The \HI\ is mainly in large tails at large radii, while
the gas in the centre, where much star formation is still occurring,
is mainly molecular.  NGC 7252 is a strong emitter in the far
infrared.  The $3^{\rm rd}$ stage with AGN activity is represented in
the sequence by B2~0648+27.  From the time-scale derived both from the
\HI\ and from the stellar population, (and given the shorter
time-scale of the AGN phase compared to mergers) the AGN activity must
occur at a late stage of the merger while a young stellar population
is still observable in the optical spectrum.  The final stage
(NGC~5266; 4\( ^{\rm th} \) panel), shows a galaxy that has become a
genuine early-type galaxy \cite{Mo1}.  In NGC~5266 the
\HI\ is falling back from the large tidal  tails to the galaxy and is in the
process of forming a large disk or ring-like structure of low surface
density.  Star formation is occurring at a much reduced rate and no
AGN is detected.  
\\
This is, of course, likely to be an oversimplified scenario as a
number of other ``free parameters'' are also playing a role. For
example, the genesis of the merger depends also on the type of
encounter (e.g. prograde vs retrograde) and the relative size of the
two galaxies is making the story more complicated. The environment is
also likely to play a very important role.

\begin{figure*}
\centerline{\psfig{figure=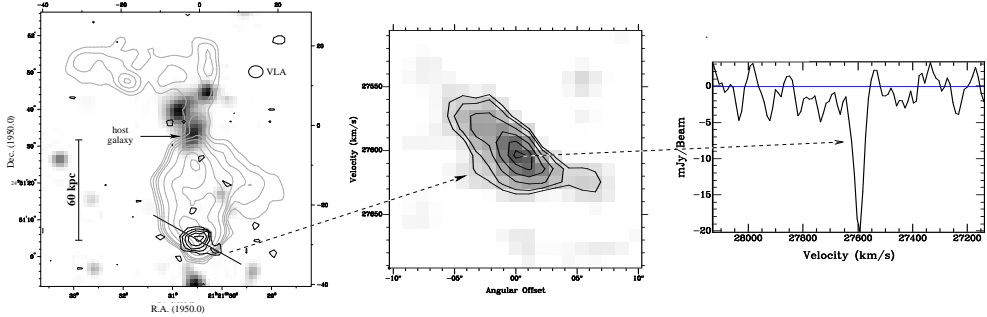,angle=-90,width=13cm}} \caption{The
preliminary total intensity image (black contours) of the \HI\ absorption in
3C433, with superimposed the  radio continuum (grey contours) and the optical
image (grey scale).  In the middle,
is shown the position-velocity plot taken along the line marked on the left
image and on the right the \HI\ profile of the peak absorption (the velocities
are in radio definition).} \label{f1} \end{figure*}

\section{Implication for nearby and high-{\it z} radio galaxies}

The initial results obtained in the study of extended structures of neutral
hydrogen in and around radio galaxies indicate that these structures may
actually be more common than thought so far.  Thus, the \HI\ can be used to
study the likely origin of these systems and the time scales involved.  Two of
the three galaxies described above show a young stellar population components. 
The parameters obtained from the study of the optical continuum and from the
\HI\ can be therefore complementary in order to obtain a full picture.  It is
important to note that galaxies with a young stellar population appear to be
the best candidates for detecting \HI.  This may further support the idea that
they come from particularly gas-rich mergers or they are mergers that occurred
recently.  \\ Finally, an important connection is between nearby and high-$z$
galaxies.  Although interactions and mergers are likely to occur more
frequently and more efficiently at high redshifts, in the \HI-rich, low
redshift radio galaxies we may witnessing similar phenomena.  Particularly
relevant is the finding (\cite{V1} and Villar-Marti\'n these Proceedings) of a
low surface brightness Ly$\alpha$ halo with quiescent kinematics in the case
of the high-$z$ radio galaxy USS~0828+193.  One possible way suggested to
explain this structure is that the low surface brightness Ly$\alpha$ halo 
is the progenitor of the \HI\ discs found in low redshift
galaxies.  \\ If this is the case, the wealth of details that we can learn at
low-$z$ may be crucial for understanding the structures at high-$z$.

\begin{figure*}
\centerline{\psfig{figure=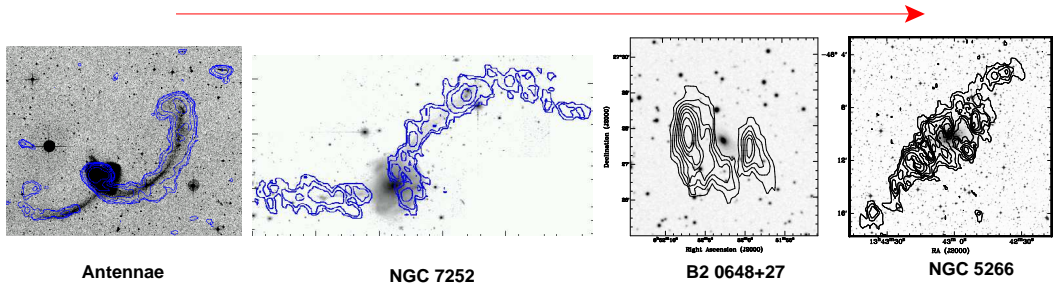,angle=0,width=14cm}}
\caption{
Possible evolutionary sequence linking gas-rich mergers with radio
galaxies and gas-rich ellipticals. The contours
represent the total \HI\ superimposed to the optical image. The
images of the Antennas and NGC~7252 have been taken from \cite{Hi2},
B2~0648+27 from \cite{Mo3}, NGC~5266 from the data presented in
\cite{Mo1}.}
\label{f1}
\end{figure*}



\begin{thebibliography}{999}

\bibitem{B1} Barnes J.E. 2002, MNRAS 333, 481

\bibitem{He1} Heckman T.M. et al. 1986 ApJ, 311, 526

\bibitem{Hi1} Hibbard J.E. 1995, PhD Thesis Columbia University

\bibitem{Hi2} Hibbard J.E. \& van Gorkom J.H. 1996, AJ 111, 655

\bibitem{Hi3} Hibbard J.E. et al., 2002 in
"Gas \& Galaxy Evolution", J.E.  Hibbard et al. (eds.), Vol. 240,
also available at http://www.nrao.edu/astrores/HIrogues/
\bibitem{Kn1} Knapp G.R., Turner E.L., Cunniffe P.E. 1985 AJ 90, 454  
\bibitem{K1} Kormendy J., Sanders D.B. 1992, ApJ 390, L53
\bibitem{Mo1} Morganti R. et al.,  1997, AJ, 113, 937

\bibitem{Mo2} Morganti R. et al.  2002, A\&A 387, 830

\bibitem{Mo3} Morganti et al. 2002, A\&A in press (astro-ph/0212323)

\bibitem{To1} Oosterloo T.A. et al.  2002, AJ 123, 729

\bibitem{To2} Oosterloo et al. 2002 in "Gas \& Galaxy Evolution", 
J.E.  Hibbard et al. (eds.), Vol.  240, p.  251

\bibitem{S1} Sadler E.M., Oosterloo T.A., Morganti R., 2002 {\em The Dynamics,
Structure \& History of Galaxies} ASP Conference Series, in press
(astro-ph/0205151)

\bibitem{Ta1} Tadhunter C.N. et al.\  2002, MNRAS 330, 977

\bibitem{V1} Villar-Marti\'n M. et al. 2002, MNRAS in press (astro-ph/0206118)

\bibitem{Wi1} Wills K.A. et al. 2002, MNRAS 333, 211


\end{thebibliography}
\end{document}